\documentclass[sigconf,printacmref=false]{acmart}
\setcopyright{none}
\renewcommand\footnotetextcopyrightpermission[1]{}
\acmConference{}{}{}  

\AtBeginDocument{%
  }

\begin{document}

\title{Adaptive LLM Agents: Toward Personalized Empathetic Care}

\author{Priyanka Singh}
\email{priyanka.singh@uni-wuerzburg.de}
\author{Sebastian Von Mammen}
\email{sebastian.von.mammen@uni-wuerzburg.de}
\affiliation{%
  \institution{University of Wuerzburg}
  \city{Wurzburg}
  \country{Germany}
}

\renewcommand{\shortauthors}{Singh et al.}


\begin{abstract}
Current mental-health conversational systems are usually based on fixed, generic dialogue patterns. This paper proposes an adaptive framework based on large language models that aims to personalize therapeutic interaction according to a user’s psychological state, quantified with the Acceptance of Illness Scale (AIS). The framework defines three specialized agents, L, M, and H, each linked to a different level of illness acceptance, and adjusts conversational behaviour over time using continuous feedback signals. The AIS-stratified architecture is treated as a diegetic prototype placed in a plausible near-future setting and examined through the method of design fiction. By embedding the architecture in narrative scenarios, the study explores how such agents might influence access to care and therapeutic relationship. The goal is to show how clinically informed personalization, technical feasibility, and speculative scenario analysis can together inform the responsible design of LLM-based companions for mental-health support.
\end{abstract}

\keywords{Conversational Agents, Mental Health, LLMs, Personalization, Illness Acceptance}

\maketitle

\section{Introduction}

Mental health is a fundamental component of human well-being, influencing how individuals think, feel, and interact with the world. It is not merely the absence of illness but a dynamic state of psychological resilience that allows a person to cope with stress and participate in daily life. When mental health is compromised, as in conditions like depression or anxiety, the impact is pervasive. These conditions arise from a combination of genetic, biological, and environmental factors, and are linked to changes in brain structure and function that alter neural circuits governing emotion and cognition~\cite{Gooding2025}.

The world is facing a mental health crisis of unprecedented scale. The World Health Organization estimates that nearly one billion people live with a mental disorder, yet a vast majority do not receive the care they need~\cite{GualMontolio2022,Jin2025}. This ``treatment gap'' is driven by systemic barriers, including high costs, a shortage of trained professionals, and persistent social stigma~\cite{Stade2023}. These barriers have motivated extensive work on digital interventions, including web-based programs, smartphone applications, and conversational agents, as potential means to augment conventional services and extend reach.

Early rule-based conversational systems showed that text-based chatbots can reduce depressive and anxious symptoms and build a perceived therapeutic alliance. With the advent of Large Language Models (LLMs), conversational agents can engage in more naturalistic and context-sensitive dialogue. However, a single, general-purpose LLM deployed as a chatbot is neither a therapist nor a clinical decision support system. Generic deployments tend to skip structured assessment, offer advice without staging, and simulate empathy without grounding it in psychotherapeutic mechanisms. Hallucinations and inconsistent adherence to safety guidelines raise additional concerns in a domain where incorrect guidance may have severe consequences.

In response to these limitations, recent work has begun to explore clinically informed conversational frameworks in which the system adapts to user readiness through dynamically reconfigurable agent behaviour. Within this line of research, the Acceptance of Illness Scale (AIS) has been proposed as a central axis for personalization, ensuring a measurable connection between psychological acceptance and interaction style. Three specialized agents---L, M, and H---represent different therapeutic strategies aligned with low, moderate, and high levels of illness acceptance, and the overall architecture is conceived as an adaptive therapeutic ecosystem rather than a monolithic model.

The present paper extends this foundation by explicitly adopting a design-fiction lens. Rather than asking only whether such a system can be built or how it might perform in a controlled trial, the AIS-stratified architecture is treated as a diegetic prototype situated in a plausible near-future setting. The focus shifts to what kinds of futures emerge once AIS-based LLM companions become mundane components of everyday life. The analysis imagines contexts in which such agents are embedded in public health infrastructures and corporate wellness programmes, and examines how design choices at the architectural level could shape dependence, autonomy, and institutional responsibility.

\begin{figure}
  \centering
  \includegraphics[width=\linewidth]{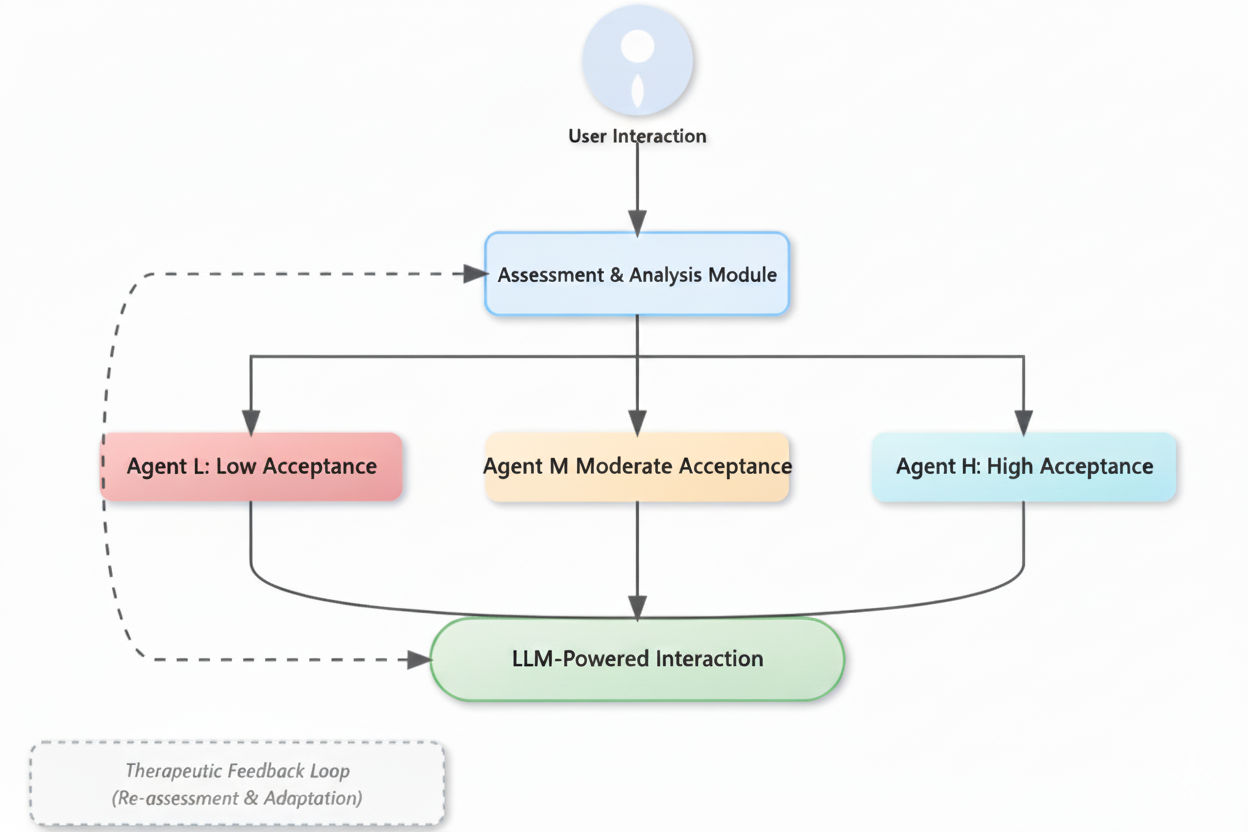}
  \caption{Therapeutic interaction model with adaptive agent acceptance. An initial AIS assessment routes the user to L, M, or H stances. Longitudinal signals feed an adaptive loop that blends agent behaviour while enforcing safety rails and supporting oversight.}
  \label{fig:therapeutic-model}
\end{figure}

The contribution of this paper is therefore twofold. First, it retains the AIS-stratified architecture with three agents (L, M, H) and an adaptive feedback loop as a technically plausible design, summarized in Table~\ref{tab:compare} and Figure~\ref{fig:therapeutic-model}. Second, it embeds this architecture into near-future scenarios as a diegetic prototype, using design fiction to illuminate advantages, risks, and societal implications that are not visible from technical evaluation alone.

\section{Background}

\subsection{Clinical grounding of AI-supported therapy}

Neuroscientific research has shown that common mental disorders such as depression and anxiety are associated with measurable changes in brain structure and function, including alterations in the hippocampus, amygdala, and prefrontal cortex, as well as dysregulated neurotransmitter systems~\cite{Gooding2025}. These changes affect the neural circuits that govern emotion, cognition, and stress regulation, and they interact with genetic, psychological, and social factors over time~\cite{Gooding2025}. From this perspective, mental health problems are not merely “feeling sad” or “being stressed”, but complex health conditions with identifiable biological and cognitive correlates.

Psychological interventions such as Cognitive Behavioral Therapy (CBT) have been developed to target these correlates at the level of thoughts, emotions, and behaviours. CBT works through mechanisms such as cognitive restructuring, behavioural activation, and guided discovery, typically delivered in a structured, staged manner. Any AI system that aims to go beyond superficial reassurance therefore needs to respect these principles: it should not simply produce supportive language, but structure interaction in a way that reflects established therapeutic processes.

\subsection{Digital interventions and the treatment gap}

The global burden of mental disorders and the lack of adequate care are well documented. Nearly one billion people worldwide are living with a mental disorder, while a large proportion do not receive appropriate support~\cite{GualMontolio2022,Jin2025}. This “treatment gap” is driven by economic constraints, shortages of trained clinicians, limited availability of services, and stigma that discourages help-seeking~\cite{Stade2023}. 

Digital interventions have been proposed as one way to partially address this gap by offering low-cost, scalable, and more accessible forms of support~\cite{GualMontolio2022}. These include web-based self-help programmes, smartphone applications, and text-based conversational agents. Early rule-based chat systems showed that structured digital programmes can reduce symptoms for some users and build a perceived sense of support~\cite{Fitzpatrick2017,Inkster2018}, but they generally rely on fixed dialogue flows and do not adapt deeply to individual differences.

\subsection{From generic chatbots to structured conversational agents}

With the advent of LLMs, conversational systems can generate more natural and context-sensitive responses~\cite{GualMontolio2022,Jin2025}. A single general-purpose LLM, prompted as a chatbot, can appear empathetic, remember previous turns within a session, and flexibly change topics. However, such generic deployments shift the burden of structure and safety onto prompting. Without an explicit therapeutic model and clear internal constraints, an LLM-based chatbot may skip assessment, move too quickly into advice-giving, or mimic empathy without a stable process. Hallucinated content and inconsistent safety behaviour add further risks in a mental health setting, where misleading or inappropriate responses can be harmful~\cite{Scholich2025,Gabriel2024}.

Recent research has therefore begun to move from “one-model chatbots” towards more structured systems in which different components are responsible for assessment, psychoeducation, change techniques, and safety monitoring. In these systems, the LLM is one element in a broader architecture rather than the entire solution. Representative multi-agent and hybrid conversational psychotherapy systems are summarised in Table~\ref{tab:compare}.

\begin{table}[t]
\centering
\caption{Comparison of multi-agent and hybrid architectures in mental health conversational systems.}
\label{tab:compare}
\small
\begin{tabular}{@{}llp{5cm}@{}}
\toprule
System & Type & Key Feature \\
\midrule
DiagGPT & Multi-Agent & Encodes protocol logic; goal-oriented dialogue \\
MedAgentSim & Multi-Agent & Simulates clinical interaction for testing \\
openCHA & Multi-Agent & Integrates reasoning and retrieval \\
ChatCounselor & Hybrid & Fine-tuned on therapist--client data \\
MindfulDiary & Hybrid & Structured journaling and reflection \\
CaiTI & Hybrid & Smart-device psychotherapy delivery \\
SoulSpeak & Hybrid & Combines short- and long-term memory via RAG \\
HabitCoach & Hybrid & CBT grounding through retrieval \\
\bottomrule
\end{tabular}
\end{table}

\subsection{Acceptance of illness as a personalization axis}

Within this broader shift, the Acceptance of Illness Scale (AIS) has been proposed as a way to anchor personalization in a clinically meaningful construct. AIS is a self-report instrument that measures how far a person has come in accepting their health condition, including perceived limitations, loss of control, and integration of the illness into daily life. Lower AIS scores are typically associated with denial, resistance, or marked distress, whereas higher scores indicate greater acceptance and readiness to engage in self-management.

In the AIS-stratified framework considered here, AIS is used as a primary axis for structuring the system’s behaviour. Instead of treating all users as equally ready for action-oriented tasks, the architecture defines three agent profiles—L, M, and H—aligned with low, moderate, and high levels of acceptance. Initial AIS scores are used to select a starting profile, and longitudinal conversational signals are then used to adjust stance cautiously over time. The goal is not to diagnose or treat illness, but to align tone, pacing, and type of support with the user’s current psychological position.

\subsection{Design fiction as a lens on future deployment}

While technical architectures and personalization schemes can be described abstractly, their implications depend strongly on how they are embedded in real-world contexts. Design fiction is a method that uses speculative but plausible near-future scenarios to explore such implications. Rather than evaluating a system only in terms of current performance metrics, design fiction situates it in concrete everyday situations and asks how it might reshape relationships, responsibilities, and power structures.

In the present work, the AIS-based, multi-agent architecture is treated as a diegetic prototype: a fictional but technically plausible system placed inside imagined near-future settings. These scenarios focus on individuals who experience symptoms of depression and anxiety but are not yet diagnosed and often do not have access to professional care. By following how such a system might be used in daily life, the analysis explores how clinically informed personalization, technical feasibility, and social context could interact, and what kinds of risks and opportunities might emerge if such companions were to become a common part of personal mental health support.

\section{Methodology}

The system follows four conceptual phases: assessment, multi-profile deployment, adaptive feedback, and situated exploration through design fiction. Assessment supplies a baseline of acceptance; multi-profile design enforces distinct therapeutic stances aligned with AIS strata; adaptation tracks longitudinal change; design-fiction scenarios embed the system into social contexts where its implications can be examined qualitatively.

At first contact, AIS is administered to estimate readiness for self-management. Scores of 8--18, 19--29, and 30--40 indicate low, moderate, and high acceptance and initialize the L, M, or H agent, respectively. This aligns the starting conversational stance with the user's psychological state and is intended to reduce early mismatch and dropout.

\subsection{Agent design, goals, and adaptation}

The L, M, and H agents share an LLM engine but differ in therapeutic focus. The L-agent prioritizes alliance and gentle psychoeducation, normalizing ambivalence and using motivational interviewing micro-skills before proposing structured tasks. The M-agent assumes a user who recognizes their condition and is willing to experiment with change, focusing on cognitive restructuring and behavioural activation tailored to the user’s context. The H-agent emphasizes resilience, relapse prevention, and adaptive self-regulation, treating the user as an active collaborator and nudging toward professional care when warning signs or chronicity patterns emerge.

Dynamic adaptation is governed by a monitored loop that tracks shifts in language, behavioural follow-through, sentiment, and explicit user feedback. When signals cross conservative thresholds, the system blends in the next stance (e.g., from pure L to mixed L--M) rather than switching abruptly. On uncertainty, it rolls back to the last verified stance. Safety rails include escalation triggers for human review and crisis protocols.

\subsection{Implementation as diegetic prototype}

Technically, the agents share a common LLM backbone and are configured through distinct system-level instruction sets and layered prompt engineering. System instructions define role, goals, forbidden content, and crisis responses. Prompt scaffolds manage local reasoning, including summarization frames, question templates, and reflective formulations. Retrieval-augmented generation grounds psychoeducation and homework structures in vetted resources, and lightweight fine-tuning may be used for format fidelity while leaving core clinical content under retrieval control.

Reasoning aids such as Chain-of-Thought and Chain-of-Empathy are used internally to guide the LLM’s reasoning and affective attunement. These chains are not exposed verbatim to the user but shape the sampling process and help maintain coherence and empathy across turns.

\subsection{Scenario construction and design-fiction lens}

To explore how the AIS-stratified companion might reshape everyday life and institutions, three narrative scenarios are constructed, set in a plausible near future where such systems are integrated into different domains: an adolescent at home (low acceptance), an employee in a corporate wellness programme (moderate acceptance), and an adult managing high acceptance in daily life. These scenarios are grounded in current practice and technical feasibility but are explicitly fictional, allowing examination of questions that cannot yet be empirically tested.

\section{Scenarios of AIS-Stratified Care Futures}

\subsection{Hypothesis 1: Low acceptance at home}

A seventeen-year-old named Riya has been withdrawing from school and social activities. Her parents encourage her to install the companion. At first contact, Riya completes the AIS questionnaire with visible reluctance and scores in the low-acceptance range; the system initializes the L-agent.

Over the following weeks, the L-agent avoids prescriptive advice and instead mirrors Riya’s feelings, normalizes ambivalence, and offers brief explanations about depression as an illness. When Riya writes that everyone expects her to “just be normal again”, the agent acknowledges the pressure she feels and validates her sense of being overwhelmed. Occasionally, it suggests small experiments, presented as options rather than obligations. The system detects repeated references to hopelessness and flags a recommendation for human follow-up, where a school counselling service sees a high-distress, low-acceptance profile and a request to consider outreach.

\subsection{Hypothesis 2: Moderate acceptance at work}

Jonas, a forty-two-year-old employee with a chronic mood disorder, accesses the AIS-stratified companion through a company wellness initiative. He receives a moderate-acceptance score, which activates the M-agent.

The M-agent collaborates with Jonas to identify stressors at work, offers cognitive reframing exercises that fit into short breaks, and proposes behavioural activation tasks around his commute. The system logs his engagement and adjusts difficulty and frequency accordingly. At the organizational level, anonymized AIS distributions and engagement statistics are displayed to human resources and management. The company uses these metrics to argue that its programme improves “resilience”, and over time a subtle expectation emerges that employees should use the platform and demonstrate proactive coping.

\subsection{Hypothesis 3: High acceptance in daily life}

Sara, a thirty-five-year-old graphic designer, has lived with recurrent episodes of depression since her early twenties and now has a relatively high level of acceptance. Her AIS score falls in the high-acceptance range, so the companion configures the H-agent as the primary stance.

The H-agent is integrated into Sara’s routines, helping her maintain a relapse-prevention plan, monitor early warning signs, and schedule restorative activities. When she notices subtle changes in sleep or concentration, she records these observations and receives suggestions grounded in past patterns and coping strategies. When warning signs accumulate over several days, the H-agent becomes more insistent about activating Sara’s safety plan, reminding her of trusted friends, peer-support options, and the possibility of professional help if symptoms worsen.

\section{Discussion and Conclusion}

Across these scenarios, the AIS-stratified companion remains technically the same system: it initializes agent stance with AIS, adapts through a monitored loop, and enforces safety rails. What changes is the institutional and social environment in which it operates. In the home setting, it mediates relationships between an adolescent, family, and school services; in the workplace, it participates in a discourse of resilience and productivity; in everyday adult life, it becomes a reflective layer woven into routines.

This variation demonstrates that architecture is not value-neutral. Stratifying users by acceptance, monitoring longitudinal signals, and automating adaptation encode assumptions about readiness, challenge, and acceptable risk. These assumptions may be clinically grounded, yet they can be reinterpreted and repurposed by institutions for aims that diverge from individual well-being. Design fiction makes these tensions visible by placing a technically plausible system into concrete, if imagined, contexts.

The paper has reframed an AIS-stratified, multi-profile LLM companion as both a clinically grounded architecture and a design-fictional artefact embedded in near-future everyday life. By staging the same technical framework in the lives of an adolescent at home, a worker in a corporate wellness programme, and an adult managing high acceptance, the analysis shows that decisions about personalization, monitoring, and adaptation are tightly coupled with questions of autonomy, responsibility, and power. Empirical work remains essential, but empirical validation without speculative exploration risks producing technically polished tools whose societal effects are poorly understood. Here, clinically grounded personalization, robust technical design, and design-fiction-based scenario analysis are treated as complementary components of responsible development for mental-health companions.

\section{Acknowledgements} 
This research was partially funded by the *Bundesministerium für Forschung, Technologie und Raumfahrt* (BMFTR) under grant number 03DPS1128A (MEDIA) and Vogel Stiftung Dr. Eckernkamp.

\bibliographystyle{ACM-Reference-Format}
\bibliography{sample-base}

@Article{GualMontolio2022,
  author    = {Gual-Montolio, Patricia and Jaén, Irene and Martínez-Borba, Verónica and Castilla, Diana and Suso-Ribera, Carlos},
  journal   = {International Journal of Environmental Research and Public Health},
  title     = {Using Artificial Intelligence to Enhance Ongoing Psychological Interventions for Emotional Problems in Real- or Close to Real-Time: A Systematic Review},
  year      = {2022},
  issn      = {1660-4601},
  month     = jun,
  number    = {13},
  pages     = {7737},
  volume    = {19},
  doi       = {10.3390/ijerph19137737},
  publisher = {MDPI AG},
}

@Article{Jin2025,
  author    = {Jin, Yu and Liu, Jiayi and Li, Pan and Wang, Baosen and Yan, Yangxinyu and Zhang, Huilin and Ni, Chenhao and Wang, Jing and Li, Yi and Bu, Yajun and Wang, Yuanyuan},
  journal   = {Journal of Medical Internet Research},
  title     = {The Applications of Large Language Models in Mental Health: Scoping Review},
  year      = {2025},
  issn      = {1438-8871},
  month     = may,
  pages     = {e69284},
  volume    = {27},
  doi       = {10.2196/69284},
  publisher = {JMIR Publications Inc.},
}

@Article{Stade2023,
  author    = {Stade, Elizabeth Cameron and Stirman, Shannon Wiltsey and Ungar, Lyle H and Boland, Cody L. and Schwartz, H. Andrew and Yaden, David Bryce and Sedoc, João and DeRubeis, Robert and Willer, Robb and Eichstaedt, johannes Christopher},
  title     = {Large language models could change the future of behavioral healthcare: A proposal for responsible development and evaluation},
  year      = {2023},
  month     = apr,
  doi       = {10.31234/osf.io/cuzvr},
  publisher = {Center for Open Science},
}

@Article{Scholich2025,
  author    = {Scholich, Till and Barr, Maya and Wiltsey Stirman, Shannon and Raj, Shriti},
  journal   = {JMIR Mental Health},
  title     = {A Comparison of Responses from Human Therapists and Large Language Model–Based Chatbots to Assess Therapeutic Communication: Mixed Methods Study},
  year      = {2025},
  issn      = {2368-7959},
  month     = may,
  pages     = {e69709},
  volume    = {12},
  doi       = {10.2196/69709},
  publisher = {JMIR Publications Inc.},
}

@Article{Gabriel2024,
  author        = {Gabriel, Saadia and Puri, Isha and Xu, Xuhai and Malgaroli, Matteo and Ghassemi, Marzyeh},
  title         = {Can AI Relate: Testing Large Language Model Response for Mental Health Support},
  year          = {2024},
  month         = may,
  archiveprefix = {arXiv},
  copyright     = {arXiv.org perpetual, non-exclusive license},
  doi           = {10.48550/ARXIV.2405.12021},
  eprint        = {2405.12021},
  primaryclass  = {cs.CL},
  publisher     = {arXiv},
}

@Article{Inkster2018,
  author    = {Inkster, Becky and Sarda, Shubhankar and Subramanian, Vinod},
  journal   = {JMIR mHealth and uHealth},
  title     = {An Empathy-Driven, Conversational Artificial Intelligence Agent (Wysa) for Digital Mental Well-Being: Real-World Data Evaluation Mixed-Methods Study},
  year      = {2018},
  issn      = {2291-5222},
  month     = nov,
  number    = {11},
  pages     = {e12106},
  volume    = {6},
  doi       = {10.2196/12106},
  publisher = {JMIR Publications Inc.},
}

@Article{Fitzpatrick2017,
  author    = {Fitzpatrick, Kathleen Kara and Darcy, Alison and Vierhile, Molly},
  journal   = {JMIR Mental Health},
  title     = {Delivering Cognitive Behavior Therapy to Young Adults With Symptoms of Depression and Anxiety Using a Fully Automated Conversational Agent (Woebot): A Randomized Controlled Trial},
  year      = {2017},
  issn      = {2368-7959},
  month     = jun,
  number    = {2},
  pages     = {e19},
  volume    = {4},
  doi       = {10.2196/mental.7785},
  publisher = {JMIR Publications Inc.},
}

@misc{Gooding2025,
  author       = {Mike A. Gooding},
  title        = {The Brain in Mental Illness},
  howpublished = {\url{https://www.webmd.com/mental-health/brain-mental-illness}},
  year         = {2025},
  month        = apr,
  note         = {[Online; accessed 24-July-2025]},
}

\appendix

\end{document}